\documentclass[a4paper,fleqn]{cas-sc}

\usepackage[numbers,sort&compress]{natbib}
\usepackage{amsmath}
\usepackage[colorinlistoftodos]{todonotes}
\usepackage{booktabs} 
\usepackage{rotating} 
\usepackage[utf8]{inputenc}
\usepackage{tikz}
\usepackage{booktabs}
\usepackage{bbding}
\usepackage{multirow}
\usepackage{tabularx}
\usepackage{multicol}
\usepackage{lipsum}
\usepackage{verbatim}
\usepackage{graphicx}
\usepackage{array}

\def\tsc#1{\csdef{#1}{\textsc{\lowercase{#1}}\xspace}}
\tsc{WGM}
\tsc{QE}
\tsc{EP}
\tsc{PMS}
\tsc{BEC}
\tsc{DE}

\begin{document}

\let\WriteBookmarks\relax
\def\floatpagepagefraction{1}
\def\textpagefraction{.001}
\shorttitle{Short Term Electricity Market Designs: Identified Challenges and Promising Solutions}
\shortauthors{L. Silva-Rodriguez et~al.}

\title [mode = title]{Short Term Electricity Market Designs: Identified Challenges and Promising Solutions}

\author[1,2,3]{Silva-Rodriguez, L.}[orcid=0000-0002-0091-6911]
\cormark[1]
\ead{lina.silvarodriguez@vito.be}
\credit{a}  
\author[1,2]{Sanjab, A.}
\credit{a}
\ead{anibal.sanjab@vito.be}
\author[3]{Fumagalli, E.}
\credit{a}
\ead{e.m.fumagalli@uu.nl}
\author[1,2]{Virag, A.}
\credit{a}
\ead{ana.virag@vito.be}
\author[3]{Gibescu, M.}
\credit{a}
\ead{m.gibescu@uu.nl}
 
\address[1]{Flemish Institute for Technological Research (VITO), Boeretang 200, Mol, Belgium}
\address[2]{EnergyVille, Thor Park 8310-8320, 3600 Genk, Belgium}
\address[3]{Copernicus Institute of Sustainable Development - Utrecht University, Princetonlaan 8a, 3584 CB Utrecht, Netherlands}

\cortext[cor1]{Corresponding author}


\begin{abstract}
The electricity market, which was initially designed for dispatchable power plants and inflexible demand, is being increasingly challenged by new trends, such as the high penetration of intermittent renewables and the transformation of the consumers’ energy space. To accommodate these new trends and improve the performance of the market, several modifications to current market designs  have been proposed in the literature. Given the vast variety of these proposals, this paper provides a comprehensive investigation of the modifications proposed in the literature as well as a detailed assessment of their suitability for improving market performance under the continuously evolving electricity landscape. To this end, first, a set of criteria for an ideal market design is proposed, and the barriers present in current market designs hindering the fulfillment of these criteria are identified. Then, the different market solutions proposed in the literature, which could potentially mitigate these barriers, are extensively explored. Finally, a taxonomy of the proposed solutions is presented, highlighting the barriers addressed by each proposal and the associated implementation challenges. The outcomes of this analysis show that even though each barrier is addressed by at least one proposed solution, no single proposal is able to address all the barriers simultaneously. In this regard, a future-proof market design must combine different elements of proposed solutions to comprehensively mitigate market barriers and overcome the identified implementation challenges. Thus, by thoroughly reviewing this rich body of literature, this paper introduces key contributions enabling the advancement of the state-of-the-art towards increasingly efficient electricity market.

\end{abstract}


\maketitle


\section{Introduction}\label{Intro}

The current electricity market setup\footnote{The current market set-up, referred to in this paper, largely focuses on electricity markets currently in place across Europe.} has been able, to some extent, to deliver competitiveness and low prices to consumers for many years. However, the growing awareness of climate change and the development of national and international agreements aiming to reduce greenhouse gas (GHG) emissions are transforming the sector significantly \cite{UnitedNations2015}. A clear example is given by the European GHG reduction goals, which aim to reduce the GHG emissions at least 40\% by 2030 compared with the emissions level in 1990 \cite{EuropeanEconomicandSocialCommittee,EuropeanCommission}. The electricity sector, which is the largest emitting sector worldwide,\footnote{The electricity sector accounted for 30\% of the global CO\textsubscript{2} emissions in 2018 \cite{IEA2019}.} is expected to play a key role in delivering a large share of savings in CO\textsubscript{2} emissions, also via the projected electrification of the heating, industrial, and transportation sectors.

In fact, the increasing share of electricity produced by variable renewable energy sources (VRES), together with the empowerment of end-users and the increased deployment of storage, are already reshaping the power system. Such evolution represents an important challenge for the wholesale electricity market originally designed for dispatchable power plants and a largely inflexible demand. This mismatch between the current market design and the emerging electricity landscape gives rise to new market inefficiencies (in an already non-ideal market), which undermines its proper operation, impacting producers and consumers alike.

To address these inefficiencies, some authors argue that electricity markets require a profound transformation \cite{Keay2016, Newbery2018} while others find that, although the current design may not be fully suitable for the emerging landscape, only minor improvements to the current design are required \cite{ENTSO-E2019, Hogan2019}. In practice, a rather large number of new market designs have been recently proposed in the literature.
While other reviews have either successfully explored the barriers to the large-scale integration of VRES \cite{Hu2018, Peng2019a}, set general principles for re-designing the electricity market \cite{Conejo2018, Newbery2018},  or reviewed specific novel proposals, such as peer-to-peer markets \cite{Sousa2019}, no previous work has, to the best of the authors' knowledge, provided a holistic and structured approach for assessing the different proposed market design solutions and explored their potential impacts on market performance. Hence, \emph{focusing on the short-term market time-frame within the context of European electricity markets},\footnote{The short-term time-frame within European markets is composed of day-ahead, intraday, and balancing markets.} the goal of this paper is threefold: (1) provide a comprehensive review of the new  market design solutions proposed in the literature, (2) assess how each proposal  addresses the sources of inefficiency in the current and future electricity markets, and (3) identify the main challenges which must be overcome to implement the proposed solutions. 

To this end, this paper starts with the identification of the criteria that the electricity market should meet (now and in the future) as well as the barriers to fulfilling these criteria. Then, a detailed investigation of the recently proposed modifications to the current market design is provided, along with an analysis of their advantages and disadvantages, which constitutes the core of the paper. These modifications, dubbed market solutions, are categorized according to the nature of the modification they proposed. Lastly, by creating a taxonomy of the identified barriers and proposed solutions, a comprehensive representation of the state-of-the-art is generated, resulting in the identification of the barriers solved by each proposal and a series of recommendations for future work directions. 

The rest of the article is organized as follows. The criteria of an efficient electricity market along with the existing barriers are provided in Section \ref{C&B}. Section \ref{OC} to Section \ref{PC} present the proposed solutions identified from the literature, as well as an assessment of their advantages and disadvantages. A comprehensive taxonomy of barriers and solutions is provided and discussed in Section \ref{Tax}. Section \ref{Conclusions} gives directions for future research and concludes the paper.

\section{Market criteria and barriers} \label{C&B}
Electricity is a complex commodity to trade and requires a wholesale market with highly specific features \cite{Kirschen2005, Laloux2013}. The growing penetration of non-dispatchable VRES, as well as the presence of new market players (on the generation and demand sides), are increasingly imposing additional challenges to an already complex market design.
Before discussing whether a design change would positively contribute to the performance of the current and future electricity market, we first identify a set of performance criteria for the wholesale electricity market. These criteria are taken from  the literature and include both the economic and technical dimensions \cite{Cramton2017, Green2005, Kirschen2005, Ito2016, Hogan2019}. They are as follows:

\begin{enumerate}
  \item The market is designed to maximize social economic welfare (SEW); 
  \item The market price provides adequate signals to market participants in the short term, which results in cost recovery and efficient consumption, as well as in the long term, resulting in efficient investments;
  \item The market is designed so that accessibility for all players is ensured, i.e., the relevant techno-economical constraints of generation, consumption, and storage are taken into consideration in a non-discriminatory manner; 
  \item The market design respects the system physical constraints and supports system security of supply;
  \item The market design supports competition among market participants;
  \item Price consistency is ensured across sequential markets, so that any deviation of the real time price from the day-ahead price is the result of uncertain factors only.

\end{enumerate}
   
Against these criteria, a number of barriers to the efficient performance of the electricity market have been identified in the literature. These can be grouped in five different, rather broad categories and are reported below (the link with the market design criteria is indicated in parentheses).  
\begin{itemize}
    \item Pricing of externalities (1, 2): An inefficient pricing mechanism for negative and positive externalities (e.g., carbon pricing and VRES subsidies) results in price distortions in the electricity market and in a reduction of the SEW \cite{Roques2017a, Hu2018, Keay2016}. 
    
    \item Pricing of electricity (1, 2): An inadequate pricing of electricity has a direct effect on security of supply and generation adequacy in the long term, resulting in a higher risk of load shedding events \cite{Hogan2017, Conejo2018}. An example of this barrier is the limited formation of scarcity prices, which are fundamental to ensuring cost recovery \cite{Roques2017a}, supporting future generation investments \cite{Pierpont2017}, and providing price signals for the demand \cite{Cramton2017}.  
    
    \item Players constraints (3): The technical characteristics of conventional market players have been explicitly taken into consideration throughout the evolution of the market design. This is not largely the case for VRES, storage technologies, and responsive demand \cite{Conejo2018}. Failure in incorporating their characteristics prevents these technologies from actively participating in the market, and results in higher balancing and reserves costs, potential load shedding events, and VRES curtailment \cite{Philipsen2019, Hu2018, Conejo2018}.  

    \item Network constraints (4): Failure to properly represent the transmission and distribution constraints of the power system in the market results in investments in suboptimal locations and additional costs associated with redispatch adjustments, VRES curtailment, and increased reliability margins \cite{IRENA2017, Newbery2018, Conejo2018, Wu2019, Gu2011, IEA2016}. This aspect is foreseen to be even more relevant in the future, as the rapid deployment of VRES is expected to increase network congestion \cite{RenewableEnergyAgency2019}. 
    
    \item Competition (5, 6): A number of market design features is known to offer the opportunity to exert market power. This market power can stem from, e.g., a limited exposure and response of participants to market signals \cite{Klessmann2008a}, an insufficient network capacity allowing market participants to elevate market prices in importing areas \cite{Borenstein2000}, and the low consistency across sequential market segments \cite{Just2015, JandeDeckerEliasDeKeyser2019, Clo2019}. 

\end{itemize}

The following four sections review the literature discussing market modifications designed to overcome the identified barriers.\footnote{The method applied to conduct the literature search is described in Appendix A.} These proposals are organized in four groups according to the nature of the modification proposed: 
(1) modifications to the organizational structure of the market design (organizational changes -- Section \ref{OC}), (2) modifications to the formulation of the market clearing procedure (market clearing changes -- Section \ref{MCC}), (3) modifications to the commodity traded in the market (commodity changes -- Section \ref{ComC}), and (4) the out-of-the-market modifications that are the result of a policy decision making process (policy changes -- Section \ref{PC}).

\section{Organizational changes} \label{OC}
A large set of proposals from the literature suggests modifying the organizational characteristics of the current market design. These characteristics include, for example, the day-ahead (DA) market time granularity, the gate closure timing (GCT), and the auction mechanism of the reserve and DA markets. 

\subsection{DA time granularity and GCT} 

The implementation of a higher time granularity for the DA market is proposed by several authors as a highly relevant improvement to the current market design \cite{ENTSO-E2019, IEA2016, Hu2018,Newbery2018, Papaefthymiou2016, Goutte2019}. The most common time granularity for DA in Europe consist of an hourly resolution [kWh/h]. However, VRES generation and demand variability within an hour can be significant \cite{Ocker2017}. This makes the current time granularity not fully suitable to reflect the real dynamics of (future) power systems \cite{Newbery2018, Philipsen2019}. The gap between the DA time granularity and the variability of supply and demand results in a higher need for intra-day (ID) trading or for reserve procurement and activation, resulting in higher balancing costs \cite{Papaefthymiou2016}. To reduce such costs, the work in \cite{Hu2018} proposes to modify the DA time granularity from 1 hour to 15 minutes while a 10 minutes granularity is proposed in \cite{Papaefthymiou2016}. The case study by \cite{Goutte2019} shows that when a higher granularity is implemented in the DA (15 minutes), the ability of flexible resources  to ramp up/down is better remunerated. The work in \cite{Deane2014} also shows that a higher granularity improves the market's ability to capture the variability of the power system and to better represent the flexibility/inflexibility of market participants. This is especially useful for markets with high penetration of VRES. Nevertheless, the benefits of this increase in granularity come at a higher computational costs and with the need to procure data at a higher resolution \cite{Philipsen2019, Deane2014}.

The modification of the GCT from one day before delivery to a time closer to delivery has also been proposed in a number of papers, such as in \cite{ENTSO-E2019, FederalMinistryforEconomicAffairsandEnergy2015, Hu2018, Papaefthymiou2016, Weber2010}. This is intended to reduce the uncertainty and the balancing costs associated with forecast errors of VRES. According to \cite{Ackermann2012, Papaefthymiou2016}, the modification in the GCT is relevant since the VRES forecasts done for the DA market (12-36h before delivery) can include significant inaccuracy. In this regard,  \cite{Hu2018} and \cite{Papaefthymiou2016} respectively propose a DA GCT of 4h and 1h before delivery. However, this shift in CGT reduces the time needed by  operators and participants to perform their required functions (e.g., clear and communicate the market results, perform reliability analyses, and take preventive actions, which are based on nominations received from market participants).

\subsection {Discrete auctions in the ID market} 

The ID market enables market players to adjust their positions from the DA market based on, e.g., more accurate forecasts. This market segment is considered of high relevance, especially for the integration of VRES, since it reduces the need for balancing services. However, in some European countries, the ID market, which is based on continuous trading and a pay-as-bid remuneration scheme, typically shows a relatively low liquidity. This low liquidity further disincentivizes participation due to high transaction costs and limited opportunities to trade \cite{Weber2010}. In addition, the work in \cite{BudishPeterCramtonJohnShim2015} concludes that the continuous trading scheme induces a so-called race for speed, which leads to arbitrage opportunities. 

To improve the performance of the ID market, the authors in \cite{Cramton2017, Weber2010} propose to shift the ID market setup into a set of discrete auctions with a pay-as-cleared mechanism, as is already the case in a number of European markets (e.g., Italy, Spain and Portugal). Such auction mechanisms prioritize the clearing of the most cost-efficient bids, contrary to the continuous bilateral trading scheme in which bids can be accepted due to their mere availability at certain times \footnote{This first-come-first-serve setup may incentivize untruthful bidding behaviour leading to a reduction of SEW as discussed in \cite{Brijs2017}.}.  The implementation of discrete auctions would also limit the advantage that some participants may have (due to the implementation of automated trading tools) and allow smaller players, which are not equipped for continuous trading, to participate in the market \cite{Neuhoff2016, Gomez2019}. Another advantage is the improvement of liquidity indicators due to the reduction in the risk perceived by the participants \cite{Weber2010}. Finally, the implementation of discrete auctions also leads to a decrease in price volatility, which further enhances participation levels \cite{Neuhoff2016}.

However, the work in \cite{Chaves-Avila2015} highlights that the achievement of high market participation in the Spanish ID market is not only attributed to the implementation of discrete auctions but to additional features (e.g., dual imbalance settlement mechanism and unit-based scheduling).\footnote{An overview of imbalance settlement mechanisms including single and dual mechanisms can be found in \cite{VanderVeen2009,VanderVeen2016, Brijs2017}} 
The main disadvantage of the implementation of this setup compared to the continuous trading approach, is the reduction in the trading flexibility for market participants, that would have to wait for the next market clearing session instead of trading immediately as in the continuous trading setup \cite{Brijs2017, Gomez2019}.

\subsection{Co-optimization of energy and reserves} \label{CER}

The procurement and activation of reserves are a fundamental responsibility of system operators to ensure a reliable power supply. When procuring reserves, the system operators contract some capacity to be able to maintain the balance (i.e. equating generation and demand while accounting for losses) during real-time operation. In the most common set up in Europe, reserve procurement and wholesale markets are organized as two separate sequential markets \cite{DeVos2019}. In this setup, the capacity that has been committed in one market (e.g., reserve capacity procurement) cannot be offered in the subsequent market (e.g., DA market) \cite{Conejo2018}. This sequential design often results in a poor price formation, since market participants implicitly include in the reserve capacity price an estimate of the opportunity costs for not participating in the DA market \cite{FlorenceSchoolofRegulation, Sores2014, Divenyi2019}.  Moreover, this sequential design increases the likelihood of an inefficient allocation of generation resources \cite{FlorenceSchoolofRegulation}.

To account for these drawbacks, several authors propose to clear energy and reserves capacity simultaneously in the same auction \cite{Abedi2020, Artac2012, Sores2014, GoroohiSardou2016, Csercsik2020, Divenyi2019, Conejo2018}, also known as co-optimization of energy and reserves, as is already the practice in a number of North-American markets. The work in \cite{Dominguez2019a}, which studies the impact of multiple reserve procurement mechanisms in Europe, shows that the co-optimization of energy and reserves schedules less reserve capacity when compared to the sequential alternative. Moreover, this approach results in a more efficient use of resources decreasing VRES curtailment levels, the likelihood of load shedding events, and the total costs \cite{Artac2012, Dominguez2019a, Abedi2020, Gonzalez2014}.  The authors in \cite{Csercsik2020, Divenyi2019, Sores2014} highlight that the co-optimization of energy and reserves enhances the trading opportunities of market participants and contributes to more efficient bidding strategies, leading to an improved price formation.

The main challenges facing the implementation of this approach in Europe are the additional computational complexity \cite{Sores2014}, data collection requirements, and organizational challenges. According to \cite{DeVos2019}, the collection of detailed per-unit data (if required) could represent an institutional challenge for the organization of the current European markets in which individual generator data is not provided to the market operator \cite{DeVos2019}. In addition, as the DA and reserve markets are most commonly cleared by two separate entities in European markets (the market operator and transmission system operator), co-optimization of energy and reserves would require  a reorganization of the roles of the operators.

\subsection{Peer-to-peer trading and local markets}

With the increased deployment of distributed energy resources (DER), several proposals have been developed to provide suitable market schemes to effectively integrate prosumers and DER into the power system. As part of these novel proposals, peer-to-peer (P2P) energy trading is often used to create new, consumer-centric marketplaces. P2P is a distributed approach in which anyone producing and/or consuming electricity (a "peer") can directly sell or buy electricity bilaterally through a trading platform or a block-chain mechanism \cite{Parag2016, VanLeeuwen2020a}. For instance, a P2P approach is proposed as a multi-bilateral economic dispatch model with product differentiation in \cite{Sorin2019}. Alternatively, the work in \cite{Morstyn2019} proposes a sequence of forward and real-time markets, based on bilateral contracts. A blockchain-based energy management platform that considers the physical, economic, and information layers of the system is proposed in \cite{VanLeeuwen2020a}. 

Another perspective is given by community-based P2P models in which the energy needs and excess of a community are managed among its members using local energy resources \cite{Parag2016, Verschae2016}. In this approach, the trading activities between peers inside the community, as well as the interface with outsiders, are managed by a central actor known as community manager \cite{Sousa2019}. A community-based P2P model -- referred to as energy collective -- composed of two levels is proposed by \cite{Moret2019}. In the first level, members of the community trade their energy internally, while in the subsequent level, the community (as a unified player) trades the excess energy in the wholesale market (e.g. DA and BA markets). Similarly, the work in \cite{Long2017} proposes a community P2P trading structure composed of three levels, which are defined based on the structure of the distribution network. According to \cite{Moret2019}, prosumers belonging to a community achieve better economic results (i.e. lower total costs) than those trading individually in a P2P model. 

A different approach for integrating DER and active prosumers is represented by local market (LM) models  \cite{Farrokhseresht2020, Jin2020, Olivella-Rosell2018, Olivella-Rosell2016}. The work in \cite{Olivella-Rosell2016, Farrokhseresht2020} proposes an LM in which the operator manages the local resources to participate in the DA market, as in \cite{Olivella-Rosell2016}, or in the DA and BA markets, as in \cite{Farrokhseresht2020}. Local flexibility markets are proposed in \cite{Jin2020} and \cite{Olivella-Rosell2018}. In these markets, the flexibility provided by local participants (usually managed by aggregators) is provided to balancing responsible parties (BRPs) to reduce imbalance \cite{Jin2020} and to DSOs for congestion management and voltage regulation needs \cite{Olivella-Rosell2018}. Alternatively, the work in \cite{AlSkaif2020} proposes a blockchain-based configuration that enables DER (more specifically, electric vehicles) to provide balancing services to the TSO. A number of additional local and P2P market designs have been proposed in the literature. An overview of these prosumer-centric models is presented in \cite{Jin2020, Sousa2019}. 

P2P and LM models are considered highly relevant for a wide adoption of DER \cite{Parag2016, Olivella-Rosell2016}. From a local perspective, P2P and LMs empower the end-users by taking into consideration their characteristics, decisions, and preferences \cite{Sorin2019,Sousa2019}. From a system perspective, the deployment of these models would boost innovation along the entire electricity value chain driving the development of new business models \cite{Sousa2019}. Notably, novel coordination schemes between grid operators, such as the ones proposed in \cite{Delnooz2019, Gerard2018,Papavasiliou2018}, would enable power systems to benefit from local distributed flexibility. Such emerging flexibility sources could help alleviate grid challenges, such as grid congestion problems, potentially resulting in the deferral of expensive grid investment projects, as discussed in \cite{ Olivella-Rosell2018, Parag2016, Sousa2019}.

The main challenge facing P2P is their scalability when considering large scale systems with a high number of participants. This is expected to induce a heavy computational burden, not only due to the  amount of participants or their different roles, but also due to the number of transactions and negotiation processes \cite{Parag2016, Sorin2019, Sousa2019}. Another challenge lies in the substantial technical and regulatory issues that must be overcome to ensure the secure operation of the power system when more local and P2P schemes are implemented \cite{Parag2016}. In addition, LM mechanisms can be impeded by low levels of consumer participation -- stemming from a low willingness to participate in complex energy decisions -- and by the risk of gamification by market participants \cite{Parag2016, Sousa2019, Sorin2019}.

\section{Market clearing changes} \label{MCC}
A number of proposals from the literature suggests modifications to the formulation of the market clearing procedure. As illustrated in this section, these proposals include smart bids, nodal pricing, and uncertainty-based market clearing models. 

\subsection{Smart orders} 

A key component of the electricity market clearing is the bid format, which is used by market participants to provide the information needed to clear the market. The traditional bid format, which is composed of a quantity [MWh] and a price [€/MWh] pair per time interval, may hinder the participation of new market players \cite{Bondy2018}. For example, the authors in \cite{Liu2015} argue that the traditional price-quantity format is too restrictive to allow the demand to fully express its flexibility.  
In addition, the traditional format does not provide the market with all the relevant (techno-economic) information to efficiently clear the market, such as ramp constraints or minimum revenue to ensure cost recovery \cite{Poli2011}.  
In light of the need to consider additional constraints, block and complex bids have been implemented in several markets around the world. For example, the European Market Coupling Algorithm (known as EUPHEMIA) includes advanced bid formats, such as minimum income condition, ramp constraints, scheduled stop, linked block orders, block orders in an exclusive group, and flexible hourly orders \cite{NEMOCommitee2019}. In addition, the work in \cite{Vlachos2013} proposes to implement block orders that can be partially cleared (known as adjustable profile blocks), These would eliminate the presence of paradoxically accepted and rejected block orders, as well as the need to use iterative heuristic procedures, which complicates and delays the market-clearing procedure.\footnote{Paradoxically accepted (rejected) bids are bids that are accepted (rejected) even if they are out(in)-the-money.}

A different angle is covered in \cite{Bondy2018, Divenyi2019, Csercsik2020}, where new bid formats are proposed for reserve capacity procurement. In this regard, the authors in \cite{Bondy2018} propose a redefinition of the requirements for the provision of reserves by taking into consideration, for example, the ramp time and the total duration of service provision. Also, the market-clearing model proposed in \cite{Divenyi2019} allows the submission of combined bids, which include in a single bid both the energy and the upward reserve offers, limited by the total generation capacity. The work in \cite{Csercsik2020} proposes to implement flexible production bids to clear a co-optimized DA and reserve procurement market. The flexible production bids take into account the ramp constraints, the start-up cost, and the variable cost. 

The works in \cite{Brolin2020, Shariat2020} propose alternative bid formats for multi-carrier energy markets\footnote{Multi-carrier markets are discussed in Section \ref{ComC}.}. The bid format proposed in \cite{Brolin2020} contains information about the energy carrier and the location of the generation unit. The authors in \cite{Shariat2020} propose conversion and storage orders. On one hand, conversion orders allow trading in one energy carrier depending on the market prices of another energy carrier. On the other hand, storage orders, allow market participants to trade energy across periods of the market clearing horizon.

Regarding demand bids, the work in \cite{Vlachos2013} proposes to implement joint block bids. This type of bids is composed of a set of demand blocks located at different nodes that should be cleared or rejected altogether. Three additional bid types are proposed in \cite{Liu2015}, to encourage higher participation from the demand and storage sides, namely, adjustable demand, deferrable demand, and arbitrage bids. The latter is proposed explicitly for storage facilities, which can profit from price discrepancies over time, and includes parameters such as storage levels, efficiency loss, and the charge and discharges rates.  

The redefinition of requirements and parameters considered in the bid format would facilitate the market access for new participants, such as demand aggregators and storage technologies, by providing a better representation of their technical characteristics and constraints \cite{Vlachos2013, Liu2015}. In turn, this would allow the market to take advantage of the flexibility and synergies of these new market players \cite{Bondy2018, Brolin2020}.

The main challenge facing the implementation of smart bids is the the complexity of the market-clearing procedure \cite{Brolin2020, Poli2011} often resulting in paradoxical clearing results  \cite{Divenyi2019, Gomez2019}. This complexity could introduce ambiguity regarding the generation of the market outcomes, which leads to lower confidence in the market procedure  and lower participation levels \cite{Gomez2019}.

\subsection {Nodal pricing}

Currently, European electricity markets commonly apply zonal pricing, which considers a simplified network with unlimited network capacity within each defined zone. By contrast, several authors propose a more granular locational pricing approach as a highly relevant improvement to the current market designs \cite{Conejo2018, Cramton2017, ENTSO-E2019, Hiroux2010, Hu2018, IEA2016, Papaefthymiou2016, Richstein}.  As of today, intra-zonal congestions are solved by Transmission System Operators (TSOs) using re-dispatch actions after the DA electricity markets are cleared. As the share of VRES increases and flexibility services emerge from the distribution side of the system, grid congestions within the zones are also expected to increase, further impacting the efficiency of the current model \cite{IEA2016, Richstein}. To avoid  \textit{ex-post} adjustments, a more accurate representation of the physical characteristics of the network is recommended \cite{Conejo2018}.\footnote{These physical characteristics may be represented through network models, such as the linearized (so-called "DC") power flow model and the non-linear AC power flow model.} Consistently, nodal pricing has emerged as a recommended solution for those regions with frequent and structural network congestion events.\footnote{Nodal pricing is commonly applied in, for example, North-American electricity markets.}  In fact, the implementation of nodal pricing achieves higher SEW when compared with other models, such as uniform and zonal pricing, especially when accounting for the cost savings stemming from a more efficient dispatch and lower electricity prices \cite{Newbery2018}. 

Although the implementation of nodal pricing is known to bring benefits also in the long-term (i.e., to guide generation investment in adequate locations), it also faces a strong regulatory and public opposition. The main concern is related to the exposure of consumers located in congested areas to higher prices (as compared to other consumers living in non-congested areas). 
For these reasons, the implementation of aggregated retail pricing regions, which average nodal prices across regions, is a commonly used option to limit consumer exposure in the US \cite{KarnstenNeuhoff2011, RenewableEnergyAgency2019}. Another general concern associated with nodal pricing is the risk of reduction in market liquidity and exertion of market power. In this regard, the implementation of compensation schemes, such as Financial Transmission Rights, have shown to encourage participation in US markets, improving liquidity indicators \cite{ KarnstenNeuhoff2011, Antonopoulos2020}. 

Finally, the work in \cite{Antonopoulos2020} discusses that the implementation of nodal pricing in Europe would require significant changes to the current market. Some of these changes are needed to allow a stronger interaction between TSOs and DSOs, the implementation of \textit{ex-ante} market power mitigation measures, and the allocation of the responsibility for the operation of the short term markets and of network operation to a single operator. Thus, the opposition to nodal pricing implementation is also partly due to the fact that moving to a nodal pricing approach requires fundamental changes to the current structure of the European electricity markets, increasing the implementations costs \cite{EuropeanCommission2016, RenewableEnergyAgency2019}. In order to avoid such radical modification, other innovative zonal models could be considered, for example, the work in \cite{Lete2020,Lete2020a} proposes a zonal market model with transmission switching, which could contribute to filling the gap between zonal and nodal pricing.

\subsection{Uncertainty-based market clearing}

The presence of high amounts of uncertainty in the power system augments the likelihood of activation of expensive fast-response units and the occurrence of load shedding events. This aspect is expected to be even more relevant in the future, due to the increasing penetration of VRES and uncertain demand. To minimize this risk, market and system operators must ensure that enough flexible generation capacity is set aside to cope with unexpected variations. However, the current deterministic approach to market clearing fails to consider the stochastic nature of these new participants, which could result in an inefficient dispatch \cite{J.MoralesA.ConejoH.MadsenP.Pinson2014}. This section explores the main techniques proposed to deal with high amounts of uncertainty via the market clearing mechanism. These techniques include stochastic programming (SP), chance constraint optimization (CCO), and robust optimization (RO). Most of the proposals included in this section co-optimize energy and reserves, continuing with the proposal described in section \ref{CER}.

Note that most of the proposed market-clearing formulations correspond to a centralized structure, similar to the approach implemented in the US, and different from the European decentralized structure. This centralized approach relies on unit commitment (UC) and economic dispatch (ED) models for power plant scheduling, where market participants provide detailed cost and technical information to the system operator. The decentralized approach, instead, relies on market bids submitted (which could, in fact, be portfolio-based rather than unit-based) to the power exchange. These bids typically do not account for detailed unit constraints, whose responsibility is given to each market participant \cite{Botterud2020}. 

\subsubsection{Stochastic programming}\label{SP}

As a means to handle uncertainty several authors have proposed to employ a two-stage SP model to clear the DA market \cite{Alvarez2019, Bjorndal2018,Bouffard2005, Kazempour2018, J.MoralesA.ConejoH.MadsenP.Pinson2014, Morales2012, Morales2014, Wong2007}. Contrary to  deterministic models where forecasts of the next day's operating conditions are only used by bidders to determine their offers, an SP model explicitly considers the possible realizations of uncertain variables in real-time within the market clearing mechanism. The two-stage SP approach aims at minimizing the expected system operation costs by considering, in a single formulation, the DA market dispatch costs, which includes both energy and reserve capacity costs (1st stage), and the expected costs of the balancing actions in real-time operation (2nd stage) \cite{J.MoralesA.ConejoH.MadsenP.Pinson2014}.\footnote{Note that not all SP formulations include reserve capacity bids as shown, for example, in \cite{Morales2012}.} The expected costs may include the costs associated with reserve activation, load shedding actions, and VRES curtailment decisions.  

In this approach, the uncertainty, which in most cases refers to VRES production, is represented by a set of scenarios, whose probability of occurrence multiplies the costs of the balancing actions to yield the expected costs for the balancing stage. Regarding the constraints, this approach includes energy balance equations for the DA market, real-time balance, bounds on the submitted bids, declaration of non-negative variables, and limits associated with load shedding, curtailment, and reserves activation. Depending on the modelling approach, transmission network capacity constraints can be omitted as in \cite{Bjorndal2018}, or included at different levels of detail (using AC power-flow models with relaxation techniques \cite{Alvarez2019} or linearized DC power flow models \cite{J.MoralesA.ConejoH.MadsenP.Pinson2014, Morales2012, Morales2014}). By anticipating, in the DA schedule, probable network congestions during real time operation, the likelihood of the occurrence of network issues can, in fact, be lowered \cite{J.MoralesA.ConejoH.MadsenP.Pinson2014, Morales2014}.

To simplify the two-stage SP problem, assumptions such as wind power production being the only source of uncertainty, a lossless transmission network and an inelastic demand, are common, in addition to the omission of inter-temporal constraints and minimum generation limits \cite{Bjorndal2018, J.MoralesA.ConejoH.MadsenP.Pinson2014, Morales2012}. A different perspective is given by formulations addressing security constraint unit commitment (SCUC) problems \cite{Bouffard2005, Wong2007} in which the uncertainty is associated with the occurrence of $n-1$ contingency events. The SCUC formulations in \cite{Bouffard2005, Wong2007} also include inter-temporal constraints, such as generation ramping limits.

Another relevant aspect is the definition of the remuneration mechanism. Both \cite{J.MoralesA.ConejoH.MadsenP.Pinson2014} and \cite{Morales2012} adopt a pricing scheme where the electricity traded in DA market is priced at the dual variable of the DA energy balance equality constraint, while the electricity to balance the system, as well as imbalances, are priced at a value proportional to the dual variable of the energy balance equation of the balancing stage.

\paragraph{Discussion:} \label{advSP}

By taking into consideration the expected uncertainty, SP provides a flexible ED/UC formulation which results in lower expected real-time operation costs. In fact, when the uncertainty is accurately represented, the SP model has been proven to perform better than the conventional ED in terms of the expected SEW \cite{Bjorndal2018, Morales2012, Morales2014}. 
In addition, \cite{Morales2012} argues that the SP model reduces market power and price volatility, by bringing together large trading volumes as a result of the co-optimization of DA and balancing markets.    

The pricing scheme proposed by \cite{Morales2012, J.MoralesA.ConejoH.MadsenP.Pinson2014} also proves to provide revenue adequacy for the market operator and cost recovery for power producers, but only in expectation. In this regard, the main disadvantage of the SP market clearing is that this approach fails to meet these characteristics for each scenario. This represents a risk for fast response producers who, in some scenarios, may be dispatched in loss-making positions -- with market prices lower than the prices submitted in their original bids, hence lowering their participation \cite{Morales2012, J.MoralesA.ConejoH.MadsenP.Pinson2014, Bjorndal2018, Morales2012}. To the best of the authors' knowledge, no work has yet proposed an SP formulation that successfully delivers revenue adequacy and cost recovery both in expectation and by scenario. A number of attempts have been made to fill this gap, such as the SP equilibrium model proposed in \cite{Kazempour2018}, which ensures cost recovery and revenue adequacy by scenario, but at the expense of very high system costs.   

Even though SP provides improved results in terms of expected economic efficiency, it has some relevant implementation limitations. For example, SP potentially introduces a high computational burden associated with the large number of scenarios required to accurately represent the uncertainty  \cite{Bertsimas2013, Jiang2012}. Another disadvantage of SP concerns the identification of scenarios, which are defined from the probability functions of stochastic generation resources. The work in \cite{Bjorndal2018} shows, through a case study, that wind power producers might be incentivized to misreport their wind probability functions in order to reduce their risk exposure and augment their expected profits. Moreover, they may fail to accurately consider the correlation between different wind power plants in different locations \cite{Morales2012}. 

\subsubsection{Chance constraint optimization}

CCO has been proposed as an alternative approach to accommodate uncertainty while avoiding the shortcomings of scenario-based SP models. In practice, the probability space of the uncertainties is added to the constraints, which limit the optimization problem's feasibility domain in such a way to guarantee that operational limits are not violated within a specified confidence level.

A key component in the CCO approach is the formulation of the chance constraints. These constraints are function of random variables, represented by their probability density function. Random variables correspond, for example, to the errors associated with VRES and load forecasting (represented as Gaussian distributions in \cite{Tang2017, Ratha2019} or as non-Gaussian correlated variables as in \cite{Wang2017}). 

The CCO approach has been proposed by the authors in \cite{Tang2017, Wang2017, Ratha2019} to address the ED problem of the DA market, where the objective function aims to minimize the generation and reserve capacity scheduling costs for the next day of operation, taking into consideration system reserve limits and transmission lines limits as chance constraints.
Moreover, a relatively higher number of papers propose to use a CCO approach to solve UC problems \cite{Dvorkin2019,Wang2012,Wang2017a,Wu2014,Wu2019, Wu2016}. For instance, \cite{Wu2014} proposes a CCO formulation to calculate the optimal hourly commitment of thermal units and the required spinning reserves to deal with load forecasting errors, power system outages, and variability of stochastic production. The constraints formulated as chance constraints ensure that generation and reserve capacity are enough to meet the demand and keep the transmission lines flow within limits. The risk indices used to set the probability associated with those constraints are the loss-of-load probability (LOLP) and the probability of transmission line overload (TLOP). As an alternative, a two-stage CCO approach is proposed in \cite{Wang2012,Wu2016,Wu2019}. In this formulation, the first stage minimizes the total UC costs, while the second seeks to minimize the costs associated with wind curtailment and load shedding events \cite{Wang2012,Wu2016} as well as incentive-based DR measures \cite{Wu2019}. In addition to LOLP and TLOP, the loss of wind probability (LOWP) is used to guarantee a utilization level of wind power. 
The formulation proposed by \cite{Dvorkin2019} also defines a pricing scheme to reward energy and reserve capacity under CCO approaches. This scheme ensures revenue adequacy and cost recovery under a number of conditions (i.e., positive commitment prices and minimum generation output for each generation unit set to zero). 

\paragraph{Discussion:} By including chance constraints pertaining to transmission lines overload, imbalances, and load shedding events, a CCO approach enables the market operator to limit the risks associated with the presence of uncertainty in real-time operation \cite{Wang2012,Wu2016, Wu2019}. In other words, CCO provides a way to balance reliability and economic efficiency \cite{Wang2017, Wu2014, Wang2017a}. For example, less restrictive chance constraints result in lower total costs in the DA market (e.g., transmission congestion costs, reserve costs), but increase the risk of violation of these constraints, as observed in the case study in \cite{Wu2014}. In addition, CCO ensures the attainment of targets such as high utilization of wind power production, while providing a reliable system operation \cite{Wang2012, Wu2019}. Those targets are relevant for the short term and long-term operation of the market. For example, a guaranteed high utilization of wind power production provides incentives for wind power investors and supports the reduction of CO\textsubscript{2} emissions \cite{Wang2012}. Lastly, when compared with scenario-based models, CCO models provide a more accurate representation of uncertainty by representing the random variables with their probability density functions, instead of using scenarios as in SP. Therefore, CCO models do not deal with the trade off between per scenario and expectation \cite{Dvorkin2019}.

The most critical disadvantage of the CCO approach is its underlying complexity. Some relevant challenges stem from the characterization of the random variables and their probability distributions, the need for a transformation of the stochastic chance constraints into a deterministic formulation, and the development of solution methods to address non-linearity and non-convexity \cite{Wang2017}. The relevance of addressing those challenges is evident, for instance, in the increasing attention given by several authors to proposing new solution algorithms for the CCO formulations and demonstrating their efficiency \cite{Wu2014, Wang2012, Wang2017, Tang2017}. Note that the complexity of the model increases with the number of constraints formulated as chance constraints, up to the point where it may not be possible to meet all constraints resulting in infeasibility or convergence problems \cite{Wang2017a, Wu2014}. Moreover, open questions, about the definition of the confidence level and the entity responsible for it, still remain to be answered.

\subsubsection{Robust Optimization}

Robustness is the key focus of what is known as the RO approach \cite{Bertsimas2013,J.MoralesA.ConejoH.MadsenP.Pinson2014,Jiang2012, Zugno2015}, which typically refers to meeting a certain requirement under worst case realization of an uncertainty. In the ED applications, RO aims at minimizing the dispatching costs, taking into consideration the worst-case realization in the balancing stage. In this approach, uncertainty is represented by an uncertainty set (instead of using scenarios as in SP). RO provides a feasible solution for any value of the uncertain parameters and an optimal result in the worst-case event \cite{Hu2016, Jiang2013, J.MoralesA.ConejoH.MadsenP.Pinson2014}. 

Only a few authors have proposed the use of RO for ED. Specifically, an adaptive robust optimization (ARO) model has been proposed in \cite{J.MoralesA.ConejoH.MadsenP.Pinson2014, Zugno2015} for markets with high penetration of wind power production.  The problem involves a three level (min-max-min) structure within two stages. The first stage seeks to minimize the total DA energy and reserve capacity costs. The second stage corresponds to the balancing stage and is represented by a max-min formulation. The maximization problem selects, within an uncertainty set, the worst-case realization of the uncertain variables (e.g., wind power production). The minimization problem represents the decisions to mitigate the impact of this worst-case realization. In other words, the costs of the balancing actions under such realization are minimized. With regard to the uncertainty set, the work in \cite{J.MoralesA.ConejoH.MadsenP.Pinson2014}  defines the uncertainty set as a group of linear inequalities, which includes the maximum deviation from the main forecast (i.e. upper and lower bounds) and the commonly used uncertainty budget, which ensures that the sum of deviations from different producers does not exceed a certain limit.\footnote{An additional constraint is included in \cite{Zugno2015} to correlate the variation of wind production in neighboring plants, which are expected to experience similar weather conditions.}  

Most authors have proposed RO models to solve UC problems \cite{Bertsimas2013,Hu2016,Jiang2012,Jiang2013,Zhao2013}. For example, \cite{Jiang2012} uses an ARO approach to account for the ramp events of wind power production, considering the worst-case scenario as the one with the highest wind power fluctuation in 24 hours. 
In this formulation, the uncertainty budget is defined to limit the number of periods in which the wind production differs from the forecast. The work in \cite{Bertsimas2013}, which also proposes an ARO approach, addresses the SCUC problem by defining the uncertainty set as the total variation from the nominal net injection in each node. This variation is considered to be generated from wind generation and real time demand variations. Alternative RO models are implemented in \cite{Hu2016,Jiang2013,Zhao2013}. A minimax regret robust (MRR) UC model is proposed by \cite{Jiang2013}. This approach aims to minimize the maximum regret of the DA decisions over all possible scenarios. In this model, regret is measured as the total cost difference between the current solution, whose uncertain realization is unknown, and the perfect information solution, which is the decision if the uncertain realization were known. A minimax variance robust (MVR) UC model is proposed in \cite{Hu2016}. In this model, the worst case is defined as the highest balancing cost deviation under uncertainty, which is defined by the authors as the largest difference between the highest and the lowest real time costs within all possible realizations in the uncertainty set.

\paragraph{Discussion:}

RO provides reliable results by minimizing the total system costs under the worst-case realization and provides feasible solutions for any realization within the uncertainty set \cite{J.MoralesA.ConejoH.MadsenP.Pinson2014, Hu2016}. This represents a potential advantage in view of practical applications, when compared with the SP, which only considers a finite set of scenarios \cite{Bertsimas2013,J.MoralesA.ConejoH.MadsenP.Pinson2014}. 
In addition, when compared with SP, RO models can better accommodate high variations of VRES, AS RO procures enough upwards capacity to deal with the worst-case realizations. 
In this sense, the likelihood of occurrence of load shedding events is minimized \cite{Bertsimas2013,Jiang2012, J.MoralesA.ConejoH.MadsenP.Pinson2014, Zugno2015}. 
Finally, RO reduces the volatility of dispatching costs by decreasing the occurrence of close to real-time emergency load shedding costs, activation of expensive reserve units, or re-dispatch actions due to transmission constraints \cite{Bertsimas2013, J.MoralesA.ConejoH.MadsenP.Pinson2014}.\footnote{The work in \cite{Hu2016}, which performs a comparative analysis between different RO models, concludes that the MVR model shows the best performance in terms of price volatility, while the lowest balancing costs are obtained using the ARO approach.}  

In general, RO shows a better real-life application perspective than other approaches (e.g., SP). The main reason is that the uncertainty set can be more easily defined from limited historical data (i.e. the mean and the range of variations around the mean) and can be adjusted according to the availability of the information \cite{Bertsimas2013, Jiang2012}. 
Another important aspect is the computational performance. RO models typically converge in few iterations, as demonstrated by the case studies in \cite{Bertsimas2013,Jiang2012}. In fact, as concluded in \cite{Zugno2015}, the problem scales well with the number of nodes in the power systems and with the number of uncertain parameters.   
 
The main disadvantage of using the RO approach is the conservativeness of the generated solutions, especially in the provision of upward reserves, which may result in significantly higher costs as compared with the current deterministic design \cite{J.MoralesA.ConejoH.MadsenP.Pinson2014,Jiang2013}. In fact, \cite{Bertsimas2013} indicates that those costs can be reduced by properly choosing the uncertainty budget, reducing then the conservativeness of the solution. However, adjusting the uncertainty budget represents a trade-off between economic performance and system reliability and, hence, must  be carefully carried out not to induce operational and reliability risks.

In addition, SP performs slightly better than RO in terms of expected total costs in the balancing stage, as shown in the comparative analysis between SP and RO in \cite{J.MoralesA.ConejoH.MadsenP.Pinson2014}. 
This is due to the conservative provision of upward reserves and to the fact that no downward reserve is scheduled in RO. As a result, the market cannot benefit from the possibility of re-dispatching downward reserves and ends up spilling wind power production instead \cite{J.MoralesA.ConejoH.MadsenP.Pinson2014,Zugno2015}.

\subsubsection{Hybrid formulations} 

In addition to SP, CCO, and RO, various other market clearing models have been proposed in the literature to deal with uncertainty, each entailing their own advantages and disadvantages. For example, the work in \cite{Zhao2013} proposes a unified two-stage stochastic-robust (SRO) UC model, which considers, in the second stage, both the expected and the worst-case generation costs, each weighted with a parameter set by the operator. The SRO model provides a more robust solution as compared with SP, and lower total costs when compared with RO. Total costs are, however, higher than those obtained when using the SP approach \cite{Zhao2013}.

A combination of chance constraint and goal programming (CCGP) is proposed in \cite{Wang2017a} for the UC problem. This joint formulation makes possible to adjust individually the risk levels of multiple chance constraints, when they lead to highly expensive solutions. In terms of economic efficiency, CCGP is considered to perform slightly better than CCO, as discussed in \cite{Wang2017a}. This is mainly due to the risk adjustment measures that lead to committing less expensive units for both energy and reserve capacity.

A bi-level optimization market clearing problem based on SP is proposed by the authors in \cite{Morales2014}. In this model, which separates conventional and stochastic dispatch in lower and upper level problems, the stochastic producers are dispatched by a central non-profit entity based on the costs of their uncertainty, while conventional producers are dispatched based on the traditional merit order (i.e. ascending order of marginal costs). According to \cite{Morales2014}, this model performs as well as SP in terms of system operation costs, and ensures cost recovery for the market participants under every scenario, contrary to SP. However, it fails to provide price-consistency \cite{Morales2017}.

A two-stage distributionally RO model, which is also a combination of SP and RO, is proposed by \cite{Wei2016}. This formulation differs from RO in the fact that the uncertainty set is composed of the probability density functions and there are no manually adjustable parameters (such as the uncertainty budget in RO). The main differences with respect to SP are that it considers the worst-case outcome and the exact probability density function is unknown (only a forecast of the mean and variance are required). Even though the model does not prove to be less conservative than RO, it has a similar computational performance while considering a full range of probabilistic information.  A similar model is used in \cite{Lu2019} in which a multi-period formulation and a more accurate framework to model the distribution density functions are proposed.

\section{Commodity changes} \label{ComC}

Besides organizational and market clearing changes, a number of innovative proposals in the literature suggest going beyond electrical energy trading by including other energy carriers in the market clearing procedure, or by trading electrical power instead of energy.  These changes to the traded commodity are discussed next.

\subsection{Multi-carrier markets}
Multi-carrier markets combine multiple energy carriers -- such as electricity, gas, and heat -- in an integrated market. These markets aim to address two main concerns. The first concern is the lack of coordination across energy systems, which may result in an inefficient usage of energy resources and forecasting errors \cite{OMalley2019, Ordoudis2019, VanStiphout2018, Zhang2018}. 
For instance, gas-fired power plants (GFPPs) bid in the DA electricity market with prices that may not reflect the price of natural gas, as the latter is traded separately in the gas market \cite{OMalley2019, Ordoudis2019}. Similarly, heating systems interact with power systems through combined heat and power units (CHP) plants, but are not organized around a market structure \cite{Zhang2018}. The second concern is the operational challenge related to the transfer of uncertainty from the demand and VRES to the gas and the heating systems through the shared components (e.g. GFPPs, CHP plants) \cite{ Chen2019, Jin2016}. In other words, multi-carrier markets enable an enhanced representation of multi-carrier technologies, such as GFPPs and CHPs, storage systems, and energy hubs \cite{Chen2019, Zhang2018}.\footnote{Energy hubs are integrated units where multiple energy carriers are conditioned, converted, or stored \cite{Mohammadi2017}.} 

Examples of models integrating electricity and gas markets are found in \cite{Ordoudis2019, OMalley2019}. A two-stage SP model, which couples the electricity and natural gas markets, is proposed in \cite{Ordoudis2019}. This model incorporates in the co-optimization of the DA and balancing markets a simplified gas network model, capturing the constraints on the flow and storage of natural gas. The work in \cite{OMalley2019} proposes, instead, to implement in the gas market contracts based on swing options. These contracts are meant to protect GFPP against gas price uncertainty and to provide valid information to form bids in the electricity market. A stochastic bi-level optimization model is used to determine the contract price while accounting for both systems.  

Another line of work discusses the integration of electricity, gas, and heating in a multi-carrier market. In this regard, \cite{VanStiphout2018} proposes to clear all markets simultaneously through a coupled, deterministic DA market model and introduces a new set of complex orders to express the relation among different carriers. The authors in \cite{Chen2019} propose an ARO model to optimize the dispatch of the three carriers in an urban energy system. An SP approach is proposed in \cite{Zhang2018} to clear an integrated energy system including storage technologies (i.e., batteries, heat storage tanks, and gas storage systems). The work in \cite{Shariat2020} proposes and compares a centralised and a decentralised multi-carrier market clearing algorithm including different types of constraints (labeled pro-rata and cumulative constraints). A different perspective is provided by the work in \cite{Brolin2020}, which proposes a centralized local multi-carrier market in which bid dependencies and simultaneous clearing is included. 

\subsubsection{Discussion}
The improved coordination of multiple energy markets maximizes the combined SEW by considering the impact of the dispatch decisions of a certain market (e.g. electricity market) on the remaining markets (e.g. heat and gas markets) \cite{Chen2019, VanStiphout2018}. Moreover, this approach effectively exploits the flexibility available in the different components of the integrated markets leading to lower balancing costs \cite{Ordoudis2019}. This higher flexibility would be fundamental to accommodate the uncertainty of the increasing penetration of VRES as discussed in \cite{Sorknes2020, VanStiphout2018, Chen2019} and illustrated in \cite{Zhang2018}. 

The main concern regarding multi-carrier markets is related to the increased complexity of the model. This complexity is evident through the challenging interpretation of results \cite{Brolin2020} and in the high computational time required to clear the market. For example, in the case study in \cite{Ordoudis2019}, the time required to solve the considered multi-carrier model is 17 times the time required to solve a model that mimics the current design. Additional concerns are related to organizational aspects, such as the coordination between system operators, the lower liquidity in non-electricity markets, and the possible rise of gaming opportunities \cite{VanStiphout2018}.

\subsection{Power based DA market}

According to \cite{Morales-Espana2017, Philipsen2019}, the coarse approximation used in the current design, which models production and consumption as energy in hourly blocks, results in costly and infeasible schedules (that are not suitable to capture ramp needs) and does not guarantee the momentary balance of supply and demand. To address this and other concerns, a power-based DA market is proposed by several authors as an alternative to the current energy-based DA market \cite{Fridgen2020, Philipsen2019, Morales-Espana2017}. 

In this regard, a first proposal implements multiple step-wise power profiles for each trading period \cite{ Fridgen2020}. According to the authors, this approach would allow market participants to trade products that better reflect their technical characteristics. Moreover, it would enable the system operator to better account for short-term fluctuations in consumption and production, as well as to identify the flexibility available in market. A second contribution proposes to implement a power based formulation with startup and shutdown trajectories that properly represent the operation of generators and make better use of their flexibility \cite{Morales-Espana2017}. A third proposed method envisions a market design where the results of the market clearing process would be twofold: a linear power trajectory for each participant, representing their electricity production and consumption at every instant, and a total price for the power produced in a certain time period \cite{Philipsen2019}. 
According to the authors, the implementation of such power-based market would result in cost efficient schedules and, in comparison with an energy-based design, lower balancing costs and wind curtailments. However, in general, these proposed changes would also impose higher requirements for the communication infrastructure as well as shorter data processing times \cite{Fridgen2020}.

\section{Policy changes}\label{PC}
The last set of proposals corresponds to the out-of-market modifications that are the result of a policy decision making process. The proposals in this category are mentioned briefly for completeness, but are not described in extensive detail. This is mainly due to the wider scope of these policy decisions, leading to the impracticality of reviewing them comprehensively within the scope of the current work due to space limitation.  

Three different policy interventions are frequently proposed in the literature. The first one focuses on the readjustment of technology-specific subsidies, to address distortions in the market price (e.g., VRES producers able to bid below marginal costs). The second one considers the adjustment of the carbon price which, under the current EU-ETS market, can be insufficient to internalize environmental externalities \cite{Cramton2017, Hu2018, Keay2016, Newbery2018, Sepulveda2018, Peng2019a}. The third one regards an adequate price formation in scarcity periods. The work in \cite{Hu2018, IEA2016} proposes to increase the price cap up to the value of loss load (VoLL). This would allow the recovery of fixed costs for power plants running only a few hours per year \cite{IEA2016}. As an alternative, administrative reserve shortage pricing can be implemented by system operators to adjust the DA price when it does not reflect the real value of flexibility \cite{Hogan2017}.

\section{Taxonomy of the market design modification proposals} \label{Tax}
\begin{figure}
    \includegraphics[scale=0.55]{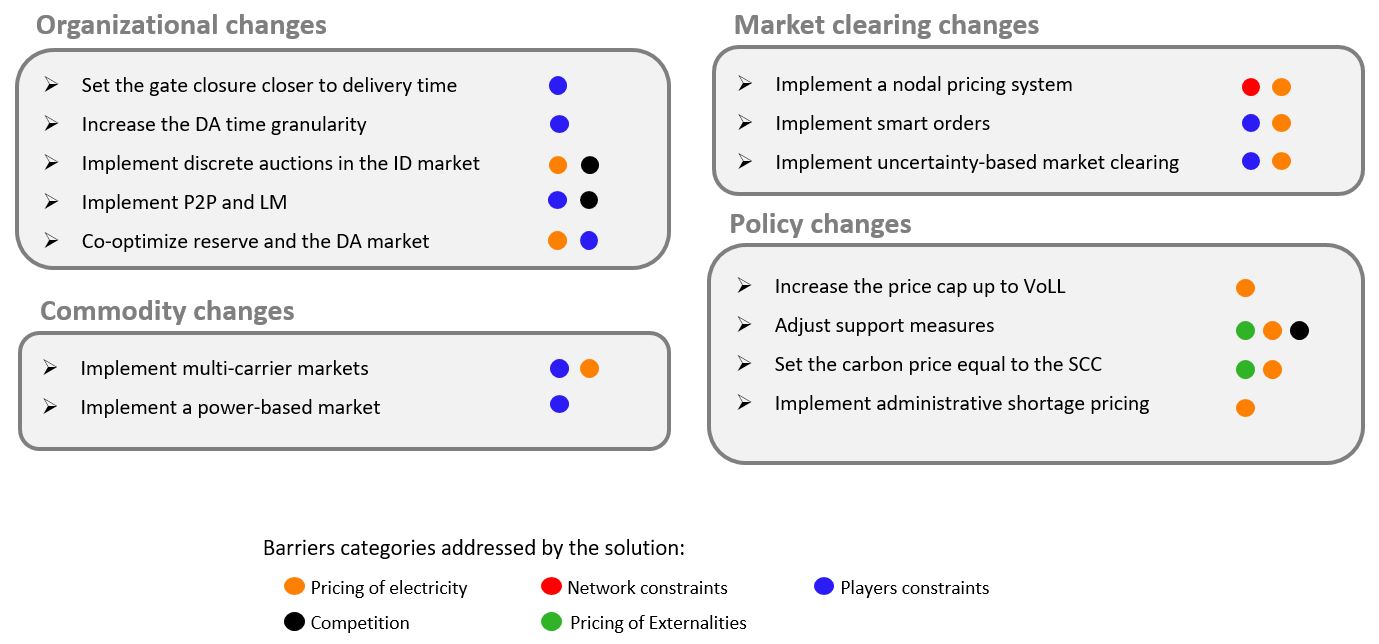}
    \caption{Barriers addressed by each proposal}
    \label{tab:Fig1}
\end{figure}

For each market design modification, Figure \ref{tab:Fig1} illustrates which of the barriers affecting the performance of the electricity market ( introduced in Section \ref{C&B}) are specifically addressed. In this regard, three main observations can be made. First, each proposal mainly focuses on one barrier category, represented by the leftmost circle. However, a proposal can address more than one barrier category simultaneously.  Second, while some barriers have received more attention than others in the recent literature, all the barrier categories have been addressed by at least one proposal for the modification of the current market design. Note, however, that when a market modification proposal addresses a barrier category, this does not imply that the proposal is capable of eliminating this barrier entirely, but rather that the proposal mitigates the effect of that barrier or partially resolves it.  
Third, although the proposals found in the literature can all be matched to one or more of the barriers, there is no proposal for a market design modification which addresses all the barriers simultaneously.

More specifically, from this taxonomy of proposed solutions and barriers, it is possible to observe the following:

\begin{itemize}
    
    \item Most of the proposals address (directly) one or more aspects of the players' constraints barrier. This highlights the focus of this wide body of literature on the presence of new players and associated generation, consumption, and storage technologies in the market. 
    \item A large share of the proposals aims at achieving a better formation of the electricity price. This goal can be interpreted in several ways and includes achieving prices that reflect the network and players' constraints (e.g., using nodal pricing and smart orders), as well as prices that fairly remunerate the availability of generation resources. 
    \item Unlike all other barriers, the network constraint barrier is effectively addressed only by one of the proposals reviewed in this paper: the implementation of nodal pricing. This could be attributed to the maturity of this solution, which is already implemented in several markets around the world (such as, in North-American electricity markets). 
    \item The modifications addressing competition issues include the implementation of discrete auctions in the ID market and the adjustment of technology-specific support measures. It should be noted that the implementation of proposals that better account for player’s constraints, such as smart orders or uncertainty-based market clearing, can also improve participation levels -- and, hence, competition -- by opening up the market to new players. 
   
\end{itemize}
 
In sum, Figure \ref{tab:Fig1} shows that there is a strong and growing interest in understanding how new players' constraints can be better represented in the short-term trading arrangements for electricity (and reserve procurement). This is not entirely surprising given their expected key role in the de-carbonization of the electricity sector. At the same time, however, there is no single way to address this major barrier category, and still no consensus on which type of modification --i.e., organizational or market clearing modifications -- would be preferable. Indeed, it appears that most of the proposed solutions, in this regard, are associated with market clearing changes, leading to modifications in the way market participants present their orders, and market operators account for players' uncertainties in the market clearing mechanism. These modifications would be accounted for differently, depending on the level of centralization of the market. A clear direction of the evolution of the market, in that regard, has still to emerge.   

Significant efforts are also being devoted to the study of how electricity prices should be defined in the future. On the one hand, local and even individual (P2P) pricing are suggested in the literature to capture the  preferences of the end users. On the other hand, the literature also suggest that wholesale market prices should represent system-wide network and reliability requirements (via nodal pricing and uncertainty-based market clearing). Even more demanding are the proposals to capture the specific features of more than one commodity within a single market price. This, as a result, introduces challenges regarding the market proposals and level of coordination within the power system, which would be necessary to purse these new directions. 

Moreover, while the concern with ensuring an adequate level of competition in the electricity market is always present in all the proposed market designs, it appears not to be the main object of the proposals. Quite simply, all design changes that might ensure a larger participation are categorized as competition enhancing and vice versa. In practice, market monitoring, together with policy changes, could be further developed to enhance competition and limit the exertion of market power.   

Finally, it is important to note that a number of challenges can hinder the practical implementation of the investigated proposals. 
The most common challenge is the high computational needs associated with the implementation of some proposals. For example, the implementation of uncertainty-based market clearing, complex smart orders, and multi-carrier markets are all associated with a high computational cost. In addition, the implementation of uncertainty-based market-clearing models has high data collection requirements, and, in some proposals, such as the implementation of SP, collecting and processing the information required for the creation of scenarios is a responsibility yet to be assigned. Another challenge lies in the possibility that the implementation of some proposals could have  negative implications on one of the other criteria. For instance, a later GCT can impact the reliable operation of the power system, and P2P models challenges the secure operation of the system. Moreover, a common challenge that could impact the implementation of the proposals is the increased complexity of the market design from the players' perspective. For example, an overly complex market design can be perceived as less transparent, which could negatively impact participation levels. 
This challenge could affect the implementation of, for example, uncertainty-based market clearing, multi-carrier markets, and power-based markets. 
Finally, the implementation of some proposals 
could unequally impact market participants. The most straightforward examples are given by the implementation of nodal pricing, which would expose consumers living in congested areas to higher prices. 

\section{Conclusions and future research outlook} \label{Conclusions}

In light of the current transformation of the electricity sector, several modifications to the current market design have been proposed in the literature. To provide a comprehensive overview of the wide set of proposals, this paper has classified the proposed solutions according to the nature of the suggested modifications (i.e., organizational, market clearing, commodity-related, and policy changes). Moreover, potential barrier categories, that hinder the ideal performance of the electricity market have been defined. This includes barriers affecting: the pricing of electricity, the pricing of externalities, network constraints, players' constraints, and market competition. Subsequently, a rigorous discussion of the advantages and disadvantages of each of the proposals was provided, along with a structured assessment of the barriers addressed by each of the proposals and the associated real-life implementation challenges. In this regard, it was observed that each barrier is addressed by at least one proposed solution, while no proposed solution is able to solve all the defined barriers. Hence, further research is needed to (1) assess which set of modifications to the current market design are required to improve the performance of the market comprehensively, and (2) overcome the identified challenges, to facilitate the implementation of the proposed market modifications and leverage their benefits.  

Future research directions should extend beyond the mere design of the market to account for the impact on the bidding behavior of market participants triggered by the implementation of the new market designs, as well as the inter-dependency between sequential markets. Indeed, by incorporating aspects related to market design, players' strategic behavior, and interdependence between markets, an analysis of any proposed market solution could provide practical and influential recommendations capable of shaping the future of electricity markets. In addition to the aspects discussed in this paper, particular attention should be given to new technological trends, which have appeared in recent years and are likely to have direct effects on electricity markets and their designs, such as power-to-molecules technologies. Among a multitude of benefits, such technologies could provide additional flexibility to the power system, allowing the large-scale integration of VRES.


\section{Acknowledgements} \label{Ack}
This work is supported by the energy transition funds project ‘EPOC 2030-2050’ organised by the Belgian FPS economy, S.M.E.s, Self-employed and Energy.

\appendix
\section{Appendix. Method}

This paper reviewed the literature proposing modifications to the current short-term market designs, while focusing on the European market context. Due to the focus on the short-term market time frame, proposals on, e.g., capacity markets, were excluded. The first stage of the review process corresponded to gathering the literature that complies with the search criteria. These criteria included a set of keywords (e.g., electricity market design, future market design, short term electricity market, among others) for papers written in English with a date of publication no older than 2010. The sources considered as relevant for this review were: books, peer-reviewed journals, conference proceedings, and scientific reports published by recognized international entities. As a result of this first stage, 143 papers were gathered.

The second stage corresponded to the analysis and selection of the literature to be included in the paper. At this stage, each reference was assigned to at least one of the identified solution categories according to the type of modification to the market that the reference proposes. In this regard, the categories created were: 1) market organizational changes, 2) market-clearing changes, 3) commodity changes, and 4) policy changes. Once classified, the relevance of each paper was assessed based on whether the authors proposed one or more particular solutions, or discussed the advantages and  disadvantages related to the implementation of the proposal. This process led to a total of 100 selected references, most of them (74\%) published between 2016 and 2020. Finally, a cross-check was performed to determine the barrier(s) that each type of market proposal was able to address. 



\bibliographystyle{model1-num-names}

\bibliography{ms}



\end{document}